\documentclass[aps,floatfix,showpacs,superscriptaddress,showkeys]{revtex4}
    
\usepackage{titlesec}
\titleformat{\section}{\large\bfseries}{\thesection}{1em}{}

\newcommand{\eq}{\begin{eqnarray}}
\newcommand{\en}{\end{eqnarray}}
\newcommand{\nn}{\nonumber\\}
\newcommand{\ed}{\end{document}}

\usepackage[english]{babel}
\usepackage{graphicx,subfigure}
\usepackage[bookmarksopen]{hyperref} 

\usepackage{amssymb,amsmath,amsfonts,latexsym,graphicx,epsfig}

\usepackage{wrapfig}
\usepackage{color}
\usepackage{upgreek}
\usepackage[amsmath,thmmarks]{ntheorem}
\usepackage{multirow} 
\usepackage{booktabs} 
\usepackage{lmodern} 
\usepackage{slashed} 
\usepackage{makeidx} 
\usepackage{fancyhdr}
\usepackage{url} 
\usepackage[title,page]{appendix} 
\usepackage{epstopdf}
\usepackage{cancel}

\usepackage{soul}

\newcommand{\sle}{\ensuremath{\slashed\epsilon}}
\newcommand{\slp}{\ensuremath{\slashed p}}

\newcommand{\eqsp}{\begin{equation}\begin{split}}
\newcommand{\speq}{\end{split}\end{equation}}

\begin{document}

\title{Hyperon forward spin polarizability $\gamma_0$ in baryon 
  chiral perturbation theory}

\author{Astrid Hiller Blin}
\affiliation{Departamento de F{\'i}sica 
  Te{\'o}rica, Universidad de Valencia and IFIC, 
  Centro Mixto Universidad de Valencia-CSIC, Institutos 
  de Investigaci{\'o}n de Paterna, Aptdo. 22085, 46071 Valencia, 
  Spain} 

\author{Thomas Gutsche}
\affiliation{
  Institut f\"ur Theoretische Physik, Universit\"at T\"ubingen,
  Kepler Center for Astro and Particle Physics, 
  Auf der Morgenstelle 14, D-72076, T\"ubingen, Germany}

\author{Tim Ledwig}
\affiliation{Departamento de F{\'i}sica 
  Te{\'o}rica, Universidad de Valencia and IFIC, 
  Centro Mixto Universidad de Valencia-CSIC, Institutos 
  de Investigaci{\'o}n de Paterna, Aptdo. 22085, 46071 Valencia, 
  Spain} 

\author{Valery E. Lyubovitskij}
\affiliation{
  Institut f\"ur Theoretische Physik, Universit\"at T\"ubingen,
  Kepler Center for Astro and Particle Physics, 
  Auf der Morgenstelle 14, D-72076, T\"ubingen, Germany}
\affiliation{ 
  Department of Physics, Tomsk State University,  
  634050 Tomsk, Russia} 
\affiliation{Mathematical Physics Department, 
  Tomsk Polytechnic University, 
  Lenin Avenue 30, 634050 Tomsk, Russia} 

\today

\begin{abstract}
  We present the calculation of the hyperon forward spin polarizability 
  $\gamma_0$ using manifestly Lorentz covariant baryon chiral perturbation 
  theory including the intermediate contribution of the spin 3/2 states.  As at 
  the considered order the extraction of $\gamma_0$ is a pure prediction of 
  chiral perturbation theory, 
  the obtained values are a good test for this theory. After including explicitly 
the decuplet states, our SU(2) results
  have a very 
  good agreement with the experimental data and we extend 
  our framework to SU(3) to give predictions to the hyperons' $\gamma_0$ values. 
Prominent are the $\Sigma^-$ and $\Xi^-$ baryons as their photon transition to the decuplet is 
forbidden in SU(3) symmetry and therefore they are not sensitive to the explicit inclusion 
of the decuplet in the theory.
\end{abstract}

\pacs{11.30.Rd,12.39.Re,13.60.Fz,14.20.Jn}
\keywords{light baryons, chiral perturbation theory, polarizabilities} 

\maketitle

\section{Introduction}\label{sintro}

From the experimental study of Compton scattering on a baryon target 
one can extract relevant information about the inner structure of 
baryons. With the help of the sum rules for integral characteristics 
of the cross sections, very important observables 
like polarizabilities can be assessed. The focus of this work is the forward 
spin polarizability $\gamma_0$, which represents the deformation of a hadron
relative to its spin axis when scattering photons in the extreme forward 
direction. It is related to the photo-absorption 
$\gamma N$ cross sections $\sigma_{3/2,1/2}$ with total helicities $3/2$ 
(for parallel photon and target helicities) 
and $1/2$ (for antiparallel photon and target helicities) via the 
sum rule in Ref.~\cite{TRHT98} 
\begin{equation}
  \gamma_0=-\frac1{4\pi^2}\int_{\omega_0}^{\infty}\mathrm{d}\omega
  \frac{\sigma_{3/2}(\omega)-\sigma_{1/2}(\omega)}{\omega^3},
\end{equation}
originally found in Ref.~\cite{GellMann:1954db}. The energy $\omega_0$ is the 
threshold for an associated neutral pion in the intermediate state. 
Experimental results for the proton $\gamma_0$ were obtained in 
Ref.~\cite{Hildebrandt:2003fm} and, more recently, in Ref.~\cite{Pasquini:2010zr}. 
Furthermore, dispersion relation studies have been 
performed in Ref.~\cite{Sandorfi:1994ku} for both nucleon spin polarizabilities.

Concerning the theoretial approach, the nucleon's structure has been thoroughly 
studied with the help of effective field theories on Compton scattering data in 
Refs.~\cite{Lensky:2009uv}, \cite{Lensky:2012ag} and \cite{Lensky:2014efa}. 
The spin-dependent piece of the amplitude $\epsilon^\mu \mathcal{M}^{\text{SD}}_{\mu\nu}
    \epsilon^{*\nu}$ attracted particular interest. 
The term proportional to $\omega^3$ ($\omega$ is the photon's energy) 
contains the whole information about $\gamma_0$, via the master formula
\begin{equation}
  \gamma_0\left[\vec{\sigma}\cdot(\vec{\epsilon}
    \times\vec{\epsilon}~^\ast)\right]=
  -\frac{\mathrm{i}}{4\pi}\frac{\partial}{\partial \omega^2}
  \frac{\epsilon^\mu \mathcal{M}^{\text{SD}}_{\mu\nu}
    \epsilon^{*\nu}}{\omega}\bigg|_{\omega=0}\,,
  \label{eqgamma0}
\end{equation}
as described in Refs.~\cite{Bernard:1995dp}, \cite{TRHT98} and \cite{VijayaKumar:2011uw}. 
Here $\vec\sigma$ is the vector of Pauli matrices, $\vec\epsilon$ and 
$\vec\epsilon~^\ast$ are the polarizations of incoming
and outgoing photons, respectively, $\alpha$ is the fine-structure 
constant and $e$ the elementary charge.

Early calculations in models of chiral perturbation theory (ChPT) that include only 
nucleonic intermediate states have been performed both in a heavy-baryon 
approach as well as in fully covariant calculations as in Ref.~\cite{Bernard:1992qa}.
In Refs.~\cite{Bernard:2012hb} and \cite{Kao:2002cp} the theory was extended such as 
to include isospin-3/2 intermediate states, namely the $\Delta(1232)$ resonance.
It was found that the inclusion of the latter state greatly improved the convergence 
between theory and empirical evidence.

When considering ChPT in SU(3) models, valuable predictions about the hyperons' 
polarizabilites can be calculated, where there are no experimental data available yet. 
First results with the help of heavy-baryon ChPT were obtained in 
Ref.~\cite{VijayaKumar:2011uw} and later improved in our work, 
Ref.~\cite{Astrid_Diplom}. The predictions are expected to be more reliable 
when using a fully covariant model and introducing the $\Delta(1232)$.

Along this line
in this work we perform a calculation of the amplitude 
$\epsilon^\mu \mathcal{M}^{\text{SD}}_{\mu\nu} \epsilon^{*\nu}$ including corrections
induced both by a fully covariant version and intermediate spin 3/2 states in a
full extension to SU(3) flavor. The leading ChPT order for the quantity $\gamma_0$ is a $p^3$ calculation. 
When including the $\Delta(1232)$ resonance, 
the study of Ref.~\cite{Bernard:2012hb} uses the so-called small scale expansion 
introduced in Refs.~\cite{Hemmert:1996xg} and \cite{Hemmert:1997ye}. We opt for a different power counting scheme, 
following Ref.~\cite{Pascalutsa:2002pi} and, 
furthermore, we introduce the couplings in a
consistent dynamics, using the full $\Delta(1232)$ propagator 
as in Refs.~\cite{Pascalutsa:1999zz}, \cite{Pascalutsa:1998pw}, 
\cite{Pascalutsa:2000kd} and \cite{Pascalutsa:2006up}. 
A study comparing different effective 
field theoretical models has been performed in Ref.~\cite{Holstein:2013kia}.

The outline of this paper is as follows. In Section~\ref{slags} we give a theoretical introduction to the appropriate 
chiral Lagrangians and the power-counting used. The kinematical considerations 
and assumptions for the calculation of $\gamma_0$ are presented in 
Section~\ref{sgam0} and the results are discussed in Section~\ref{sres}. Finally, we briefly summarize in Section~\ref{summ}.

\section{ChPT involving pseudoscalar mesons, baryons and photons} 
\label{slags} 

For the description of hyperon polarizabilities we use 
a manifestly Lorentz covariant SU(3) version of chiral perturbation 
theory involving pseudosclar mesons, baryons and photons 
(see details in Refs.~\cite{SW79} and \cite{Gasser:1987rb}). 
The lowest-order chiral Lagrangian involving 
pseudoscalar mesons $\phi$, baryons $B$ and 
photons $A_\mu$ reads 
\begin{equation} \label{Lagr_full} 
  \mathcal{L} = \mathcal{L}^{(2)}_{\phi\phi} + \mathcal{L}^{(1)}_{\phi B},
\end{equation}
where 
\begin{equation}
  \mathcal{L}^{(2)}_{\phi\phi}
  =\frac{F_0^2}4\text{Tr}\left(u_\mu u^\mu +\chi_+\right)
  \label{eqmesonlag}
\end{equation} 
is the $\mathcal O(p^2)$ meson Lagrangian and 
\begin{equation}
  \mathcal{L}^{(1)}_{\phi B}
  =\text{Tr}\left(\bar{B}(\mathrm{i}\slashed{\mathrm{D}}-m)B\right)+
  \frac D2\text{Tr}\left(\bar{B}\gamma^\mu\gamma_5\left\{u_\mu,B\right\}\right)+
  \frac F2\text{Tr}\left(\bar{B}\gamma^\mu\gamma_5\left[u_\mu,B\right]\right)
  \label{eqlagmesbar}
\end{equation}
is the $\mathcal O(p^1)$ Lagrangian including baryons. 
The symbols $[ \,\, ]$ and  $\{ \,\, \}$
occurring in Eq.~\ref{eqlagmesbar} and in the following 
denote the commutator and anticommutator in flavour space, respectively. 
The vielbein $u_\mu$ is given by
$\mathrm{i}\left\{u^\dagger,\nabla_{\mu} u\right\}$, 
with $u^2=U=\exp\left(\frac{\mathrm{i}\phi}{F_0}\right)$, where 
$\nabla_{\mu} u = \partial_\mu u - \mathrm{i}(v_\mu + a_\mu)u 
+ \mathrm{i}u(v_\mu - a_\mu)$ 
and $\mathrm{D}_\mu B = \partial_\mu B + [\Gamma_\mu,B]$ are the covariant 
derivatives acting on meson and baryon octet fields, respectively, and 
$m$ denotes the baryon octet mass in the chiral limit. The chiral connection is 
given by $\Gamma_\mu = \frac12[u^\dagger,\partial_\mu u] 
- \frac{\mathrm{i}}{2}u^\dagger(v_\mu + a_\mu)u 
- \frac{\mathrm{i}}{2}u(v_\mu - a_\mu)u^\dagger$. 
Since we are working with photon fields, we set $v_\mu$ to $e\epsilon_\mu Q$ and 
$a_\mu$ to $0$. The constant $F_0$ is the meson decay constant in the chiral 
limit and the low-energy constants $D$ and $F$ are determined from hyperon 
$\beta$ decays, where the combination $F+D$ corresponds to the low-energy constant $g_A$ in the SU(2) limit. 
The explicit form of the $3 \times 3$ charge matrix $Q$, meson $\phi$ 
and baryon $B$ matrices is given in Appendix~\ref{Appendix:ChPT}. 
The term $\text{Tr}(\chi_+)$ is responsible for the explicit breaking 
of the chiral symmetry due to the finite quark masses 
\begin{equation}
  \text{Tr}(\chi_+)=\text{Tr}\left(\chi U^\dagger + U\chi^\dagger\right),
  \label{eqlagmes}
\end{equation}
where in our case $\chi=M^2$ and $M$ is the meson mass. The power-counting 
scheme followed here gives the order
\begin{equation}
  N=4N_L + \sum_{d=1}^\infty{d N_{d}} - 2P_\phi-P_B\label{eqcountbar}
\end{equation}
to a diagram, where $N_L$ stands for the number of loops, 
$N_d$ for the number of vertices 
from Lagrangians of order $d$, and $P_\phi$ and $P_B$ 
for the number of meson and baryon propagators, respectively 
(see also Ref.~\cite{Weinberg:1991um}).

In this work we also include isospin-$3/2$ resonances, 
which give significant corrections to the full amplitude. 
The relevant terms of the Lagrangians that couple these decuplet 
fields to the octets of baryons and mesons are partly given 
in Refs.~\cite{Geng:2009hh}, \cite{Geng:2009ys}, \cite{Ledwig:2014rfa}, where the kinetic term reads
\begin{equation}
  \mathcal{L}^{(1)}_{\Delta} = \bar{\Delta}_\mu^{abc}\left[
    \gamma^{\mu\nu\alpha}\mathrm{i}\partial_\alpha -M_\Delta\gamma^{\mu\nu}
    \right]\Delta_\nu^{abc}. 
\end{equation}
We added the missing couplings by 
extending the known vertices from SU(2) 
(see Refs.~\cite{Pascalutsa:2005ts} and \cite{Pascalutsa:2006up}) to SU(3):
\begin{eqnarray}\label{EqDelPhiB}
  \mathcal{L}^ {(1)}_{\Delta\phi B}&=&
  \frac{-\mathrm{i}\sqrt{2}\mathcal{C}}{F_0M_\Delta}\bar{B}^{ab}
  \varepsilon^{cda}\gamma^{\mu\nu\lambda}
  (\partial_\mu\Delta_\nu)^{dbe}
  (\mathrm{D}_\lambda\phi)^{ce} + \text{H.c.}\,,\\
  \mathcal{L}^{(2)}_{\Delta B}&=&-\frac{3\mathrm{i}eg_M}
          {\sqrt2m(m+M_\Delta)}
          \bar{B}^{ab}\varepsilon^{cda}Q^{ce}
          (\partial_\mu\Delta_\nu)^{dbe}\tilde{F}^ {\mu\nu} 
          + \text{H.c.}\label{eqlagdelt}\,, 
\end{eqnarray}
where $\Delta^{ijk}$ are the decuplet states (see 
details in Appendix~\ref{Appendix:ChPT}), $\tilde{F}^{\mu\nu} 
=\varepsilon^{\mu\nu\alpha\beta} \partial_{\alpha}A_{\beta}$ is the self-dual 
stress tensor of the electromagnetic field, and the Dirac tensors 
$\gamma^{\mu\nu}$ and $\gamma^{\mu\nu\lambda}$ are specified 
in Appendix~\ref{Appendix:ChPT}.

The couplings 
$\mathcal{C}$ and $g_M$ are low-energy constants and 
$M_\Delta$ corresponds to the decuplet mass in the chiral limit. 
Note that the constant $\mathcal{C}$ corresponds to the low-energy 
constant $h_A$ of SU(2) in Refs.~\cite{Pascalutsa:2005ts} and \cite{Pascalutsa:2006up} 
by the conversion $\mathcal{C}=-\frac{h_A}{2\sqrt{2}}$. 
This definition of $h_A$ differs by a factor $2$ from the definition found in 
Ref.~\cite{Bernard:2012hb}. This low-energy constant is extracted from the strong decay 
of the decuplet into the baryon octet and has been determined to be $h_A=2.85$ in Ref.~\cite{Pascalutsa:2004je} 
for SU(2) and $\mathcal{C}=-0.85$ in Ref.~\cite{Ledwig:2014rfa} for SU(3). It is also  
important to mention that the numerical value for the coupling constant $g_M$ 
has not yet been studied when extending the 
model to SU(3). Therefore the quality of the predictions very much depends 
on its correct value. We follow the method 
of Ref.~\cite{Bernard:2012hb} and estimate the value of $g_M$ by calculating the width 
of the electromagnetic decay of the $\Delta(1232)$:
\begin{align}\label{EqGM}
\Gamma_{\Delta}^\text{EM}=-2\text{Im}(\Sigma_{\Delta}^\text{EM})=
\frac{e^2g_M^2(M_\Delta-m)^3(M_\Delta+m)^3}{4M_\Delta^3m^2\pi},
\end{align} 
where $\text{Im}(\Sigma_{\Delta}^\text{EM})$ is the imaginary part 
of the electromagnetic $\Delta(1232)$ self-energy amplitude. Therefore, using 
the relation $\Gamma_{\Delta}^\text{EM}/(\Gamma_{\Delta}^\text{EM}+\Gamma_{\Delta}^\text{Strong})=0.55\%...0.65\%$ 
and the strong decay width $\Gamma_{\Delta}^\text{Strong}=(118\pm2)\text{MeV}$, 
we get the value $g_M=3.16\pm0.16$. Since data on the electromagnetic decays of the full decuplet are 
sparse and contain large errors, a determination of $g_M$ in the SU(3) version is not viable. We therefore 
also fix $g_M$ in SU(3) to the $\Delta\longrightarrow\gamma N$ decay, i.e we also use the value of $g_M=3.16\pm0.16$ here. 
We should keep in mind that the 
central value of $g_M$ will change when going from SU(2) to SU(3), but we expect that with the present 
sizable error on $g_M$ this value is included.

For the covariant derivative we use
\begin{equation} 
  (\mathrm{D}_\lambda\phi)^{ab} = 
  \frac{1}{\sqrt2}\left(\partial_\lambda\phi^{ab} 
  - \mathrm{i}eA_\lambda[Q,\phi]^{ab}\right).
\end{equation}
The factor $\frac1{\sqrt2}$ comes from the definition of the meson-octet matrix. 
When introducing the decuplet fields into the chiral theory, 
an additional small parameter of the ChPT expansion appears, $\delta= M_\Delta - m\sim 300\text{MeV}$. 
Therefore, the counting scheme has to be revised. 
Here we follow the $\delta$ counting scheme from Ref.~\cite{Pascalutsa:2002pi}, 
where $\delta^2$ is counted as $\mathcal{O}(p)$. 
It is adequate for the low-energy range close to pion-production threshold. 
Hence, one obtains the rule
\begin{equation}
  N=4N_L + \sum_{d=1}^\infty{d N_{d}} - 2P_\phi-P_B-\frac{1}{2}P_\Delta,
\end{equation}
where now $P_\Delta$ is the number of $\Delta$ propagators. Since the $\Delta(1232)$ is a 
spin-$3/2$ resonance, it does not have the normal Dirac-propagator form for spin-$1/2$ particles, 
but takes the Rarita-Schwinger form given by
\begin{eqnarray}
  S_\Delta^{\alpha\beta}(p) &=& 
  \frac{\slp+M_\Delta}{p^2-M_\Delta^2+\mathrm{i}\varepsilon}
  \, \biggl[
    -g^{\alpha\beta}
    + \frac{1}{d-1}\gamma^\alpha\gamma^\beta\nonumber\\
    &+& \frac{1}{(d-1) M_\Delta}(\gamma^\alpha p^\beta - \gamma^\beta p^\alpha)
    + \frac{d-2}{(d-1) M_\Delta^2}p^\alpha p^\beta
    \biggr],
\end{eqnarray}
where $d$ is the number of dimensions of the Minkowski 
space, which after dimensional regularization is set to $d=4$.

The calculations in this work are done up to order $p^3$ for 
the isospin-$1/2$ counting scheme, which is the leading-order 
calculation for the $\gamma_0$ observable, and up to order $p^{7/2}$ 
in the isospin-$3/2$ counting scheme. This choice is due 
to the fact that in the isospin-$1/2$ sector the first contributions 
appear at loop level, which corresponds to the order $p^3$ if all the 
coupling vertices are extracted from the lowest-order Lagrangian. 
Instead of going to higher-order couplings and therefore obtaining 
loops of order $p^4$, we included the isospin-3/2 sector, 
which due to its lower order --- the leading-order diagrams are at order $p^{7/2}$ --- 
is expected to dominate over those contributions. 
In addition there is the advantage that fewer additional low-energy constants are 
needed than for the case of higher orders. All the 
constants' values chosen for this work are given in Table~\ref{tabconsts}. In order to obtain the SU(2) results, 
we simply put all the channels with non-vanishing strangeness to 0 and keep only those channels involving nucleons, 
pions and the isospin-$3/2$ quadruplet.

\begin{table}[h]
  \begin{center}
    \begin{tabular}{cccccccccccccc}
      \hline \hline
                    &            & $m$  & $M_\Delta$ & $M_\pi$ & $M_K$ & $M_\eta$ & $F_0$   & $g_A$  & $D$    & $F$    & $\mathcal{C}$ & $h_A$  & $g_M$\\
      SU(2)& chiral limit choice & $880$  & $1152$   & $140$  & $--$   & $--$   & $87$  & $1.27$ & $--$   & $--$   & $--$          & $2.85$ & $3.16$\\
           & physical choice     & $938.9$  & $1232$   & $138.04$  & $--$   & $--$   & $92.21$  & $1.27$ & $--$   & $--$   & $--$          & $2.85$ & $3.16$\\
      SU(3)& chiral limit choice & $880$  & $1152$   & $140$  & $496$ & $547$   & $87$  & $--$   & $0.623$ & $0.441$ & $-D$          & $--$   & $3.16$\\
      & physical choice          & $1149$ & $1381$   & $140$  & $496$ & $547$   & $108$ & $--$   & $0.8$ & $0.47$ & $-0.85$       & $--$   & $3.16$\\
      \hline \hline
    \end{tabular}
  \end{center}
  \caption{Numerical values for the hadron masses and decay constants 
    used in the $\gamma_0$ calculations. All the dimensionfull values are given 
    in units of MeV. The physical choice values for SU(2) were taken as in Ref.~\cite{Bernard:2012hb} --- notice the 
difference by a factor 2 in the Lagrangian definitions of $h_A$ ---, 
    whereas for the chiral limit and SU(3) we followed Ref.~\cite{Ledwig:2014rfa}. The value for the coupling $g_M$
	 is calculated through Eq.~\ref{EqGM}.
  }
  \label{tabconsts}
\end{table}

\section{The forward spin polarizability $\gamma_0$}\label{sgam0}

\begin{figure}
  \begin{center}
    \subfigure[]{
      \label{vertvert}
      \includegraphics[width=0.3\textwidth]{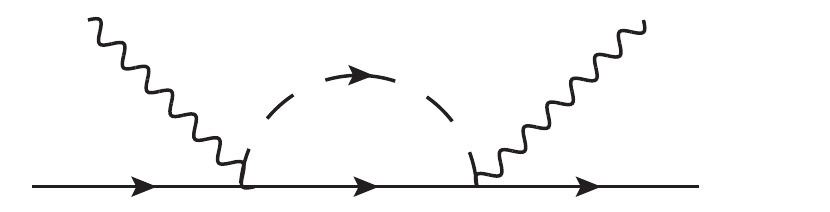}}
      \subfigure[]{
      \label{barovert}
      \includegraphics[width=0.3\textwidth]{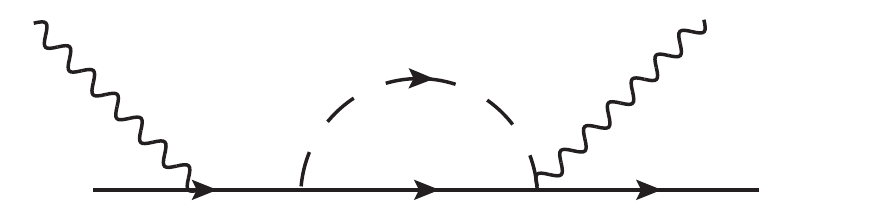}}
      \subfigure[]{
      \label{vertobar}
      \includegraphics[width=0.3\textwidth]{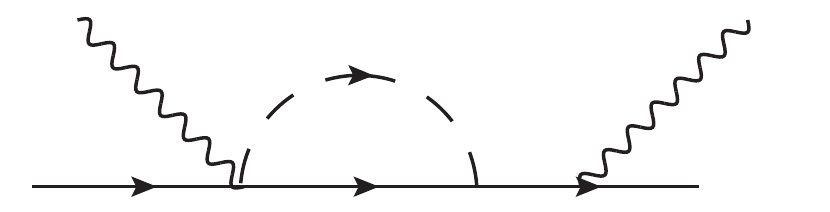}}\\
      \subfigure[]{
      \label{barsvert}
      \includegraphics[width=0.3\textwidth]{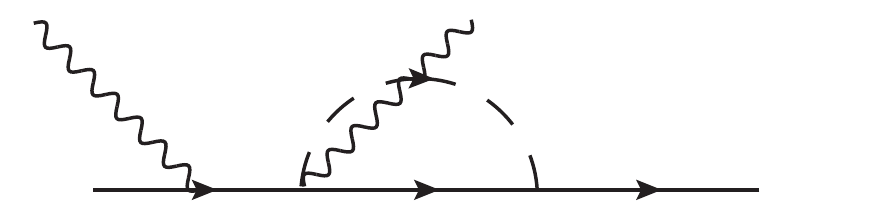}}
      \subfigure[]{
      \label{vertsbar}
      \includegraphics[width=0.3\textwidth]{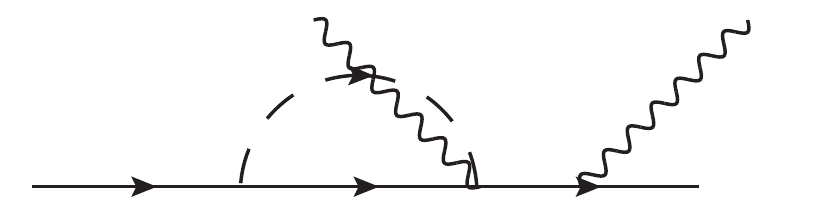}}
      \subfigure[]{
      \label{barbar}
      \includegraphics[width=0.3\textwidth]{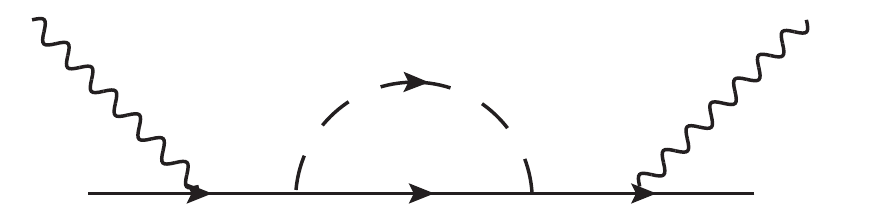}}\\
      \subfigure[]{
      \label{barmes}
      \includegraphics[width=0.3\textwidth]{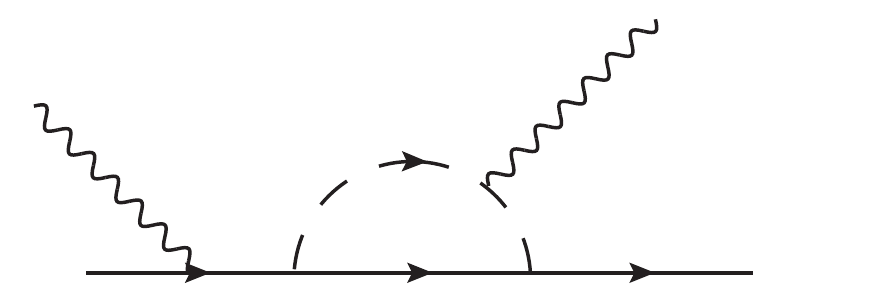}}
      \subfigure[]{
      \label{mesbar}
      \includegraphics[width=0.3\textwidth]{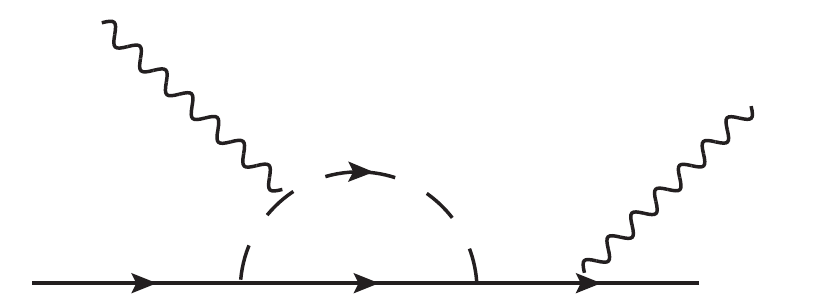}}
      \subfigure[]{
      \label{vertmes}
      \includegraphics[width=0.3\textwidth]{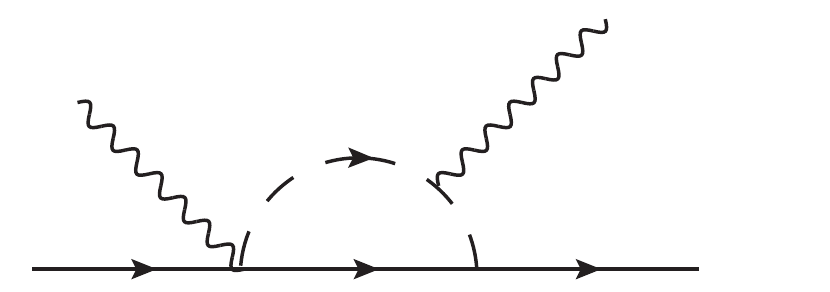}}\\
      \subfigure[]{
      \label{mesvert}
      \includegraphics[width=0.3\textwidth]{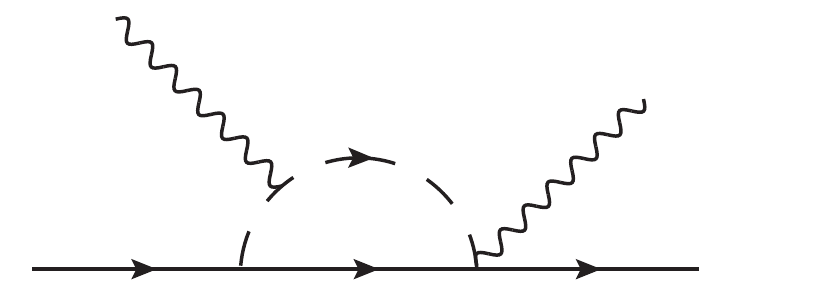}}
      \subfigure[]{
      \label{vertvirbar}
      \includegraphics[width=0.3\textwidth]{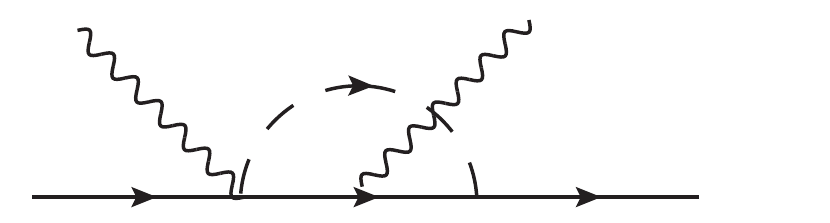}}
      \subfigure[]{
      \label{virbarvert}
      \includegraphics[width=0.3\textwidth]{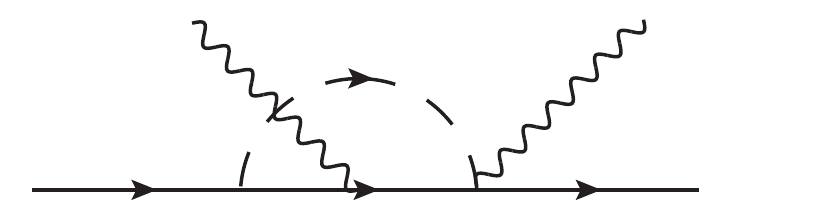}}\\
      \subfigure[]{
      \label{barvirbar}
      \includegraphics[width=0.3\textwidth]{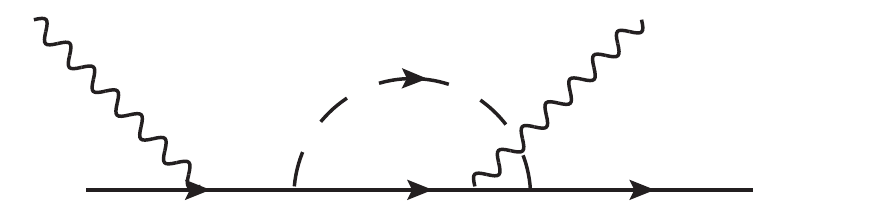}}
      \subfigure[]{
      \label{virbarbar}
      \includegraphics[width=0.3\textwidth]{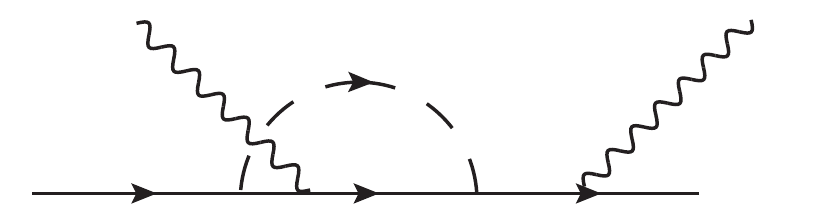}}\\
      \subfigure[]{
      \label{virbarvirbar}
      \includegraphics[width=0.3\textwidth]{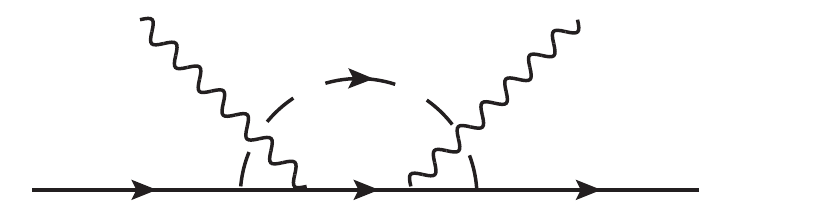}}
      \subfigure[]{
      \label{virbarmes}
      \includegraphics[width=0.3\textwidth]{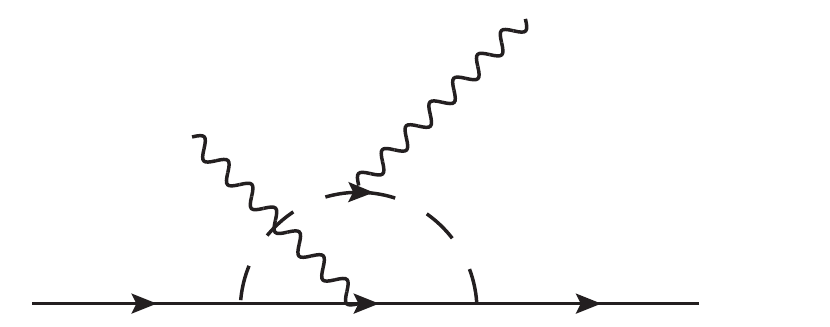}}
  \end{center}
  \caption{Diagrams contributing to $\gamma_0$ with isospin-$1/2$ 
    intermediate states. The crossed diagrams are obtained by the 
    substitutions $\omega\leftrightarrow-\omega$ and 
    $\sle\leftrightarrow\sle^\ast$.}
  \label{fiso12}
\end{figure}

\begin{figure}
  \begin{center}
    \subfigure[]{
      \label{fdeltree}
      \includegraphics[width=0.3\textwidth]{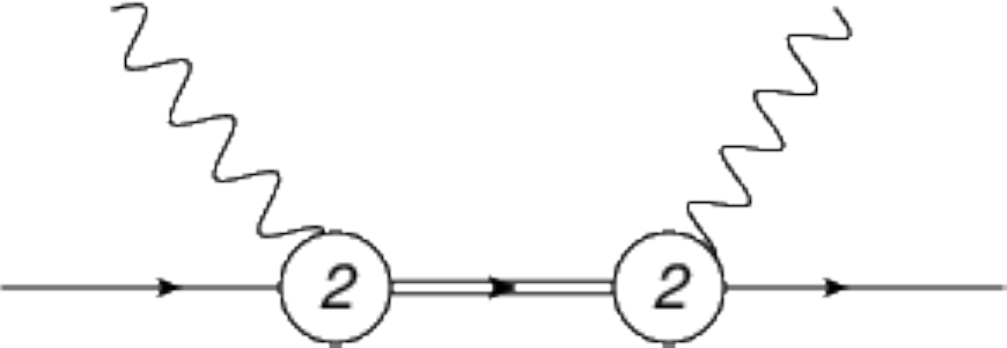}}
    \subfigure[]{
      \label{fdela}
      \includegraphics[width=0.3\textwidth]{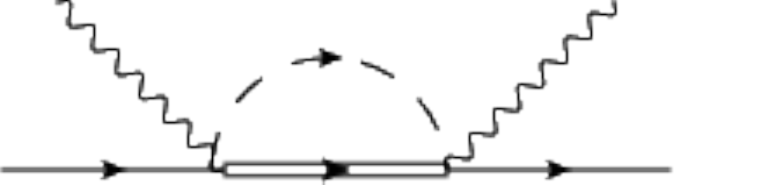}}
    \subfigure[]{
      \label{fdelb}
      \includegraphics[width=0.3\textwidth]{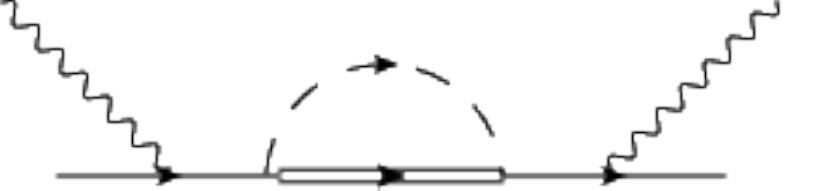}}\\
    \subfigure[]{
      \label{fdelc}
      \includegraphics[width=0.3\textwidth]{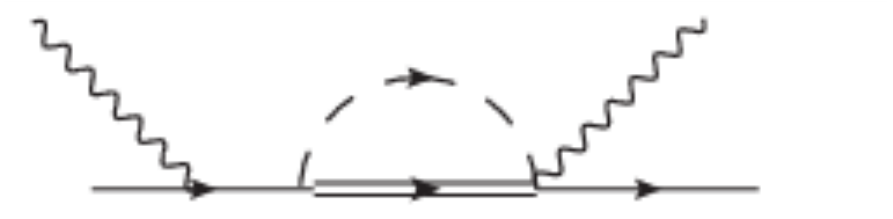}}
    \subfigure[]{
      \label{fdeld}
      \includegraphics[width=0.3\textwidth]{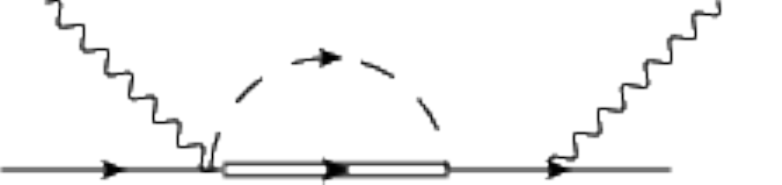}}
    \subfigure[]{
      \label{fdele}
      \includegraphics[width=0.3\textwidth]{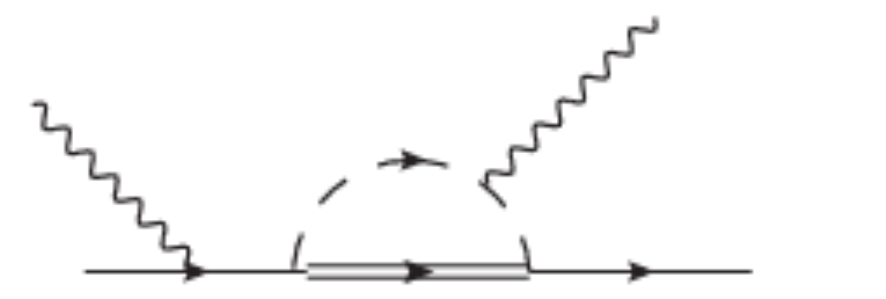}}\\
    \subfigure[]{
      \label{fdelf}
      \includegraphics[width=0.3\textwidth]{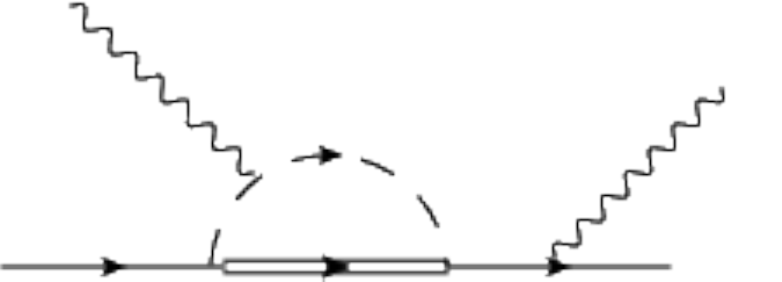}}
    \subfigure[]{
      \label{fdelg}
      \includegraphics[width=0.3\textwidth]{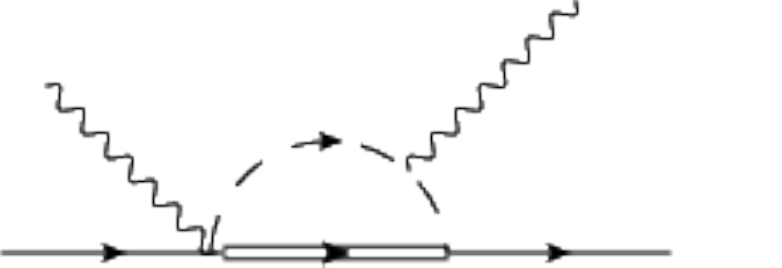}}
    \subfigure[]{
      \label{fdelh}
      \includegraphics[width=0.3\textwidth]{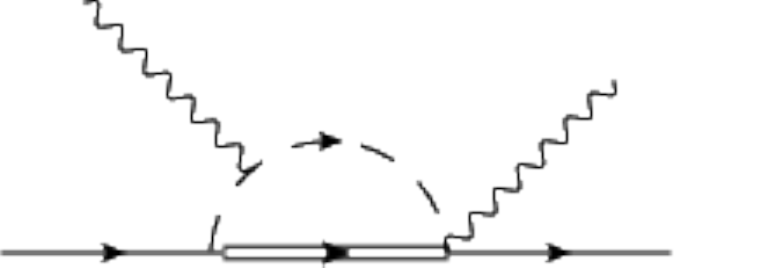}}
  \end{center}
  \caption{Diagrams contributing to $\gamma_0$ with isospin-$3/2$ 
    intermediate states. The crossed diagrams are obtained by the 
    substitutions $\omega\leftrightarrow-\omega$ and
    $\sle\leftrightarrow\sle^\ast$. Except for the tree diagram, 
    which has vertices of a second-order Lagrangian, 
    all the vertices that appear are couplings of lowest-order Lagrangians.}
  \label{fiso32}
\end{figure}

The diagrams contributing to the forward spin polarizability 
$\gamma_0$ are shown in Figs.~\ref{fiso12} and \ref{fiso32}. 
The calculation of the amplitudes corresponding to each of these diagrams is 
performed in the rest frame of the baryon. 
To calculate the forward spin polarizability one needs to assume 
conservation of the photon energy $\omega=\omega'$ ---  
incoming and outgoing photons have the same momenta $\vec q=\vec q^{\,'}$. In the following, 
the Minkowski-space vectors used are:
\begin{equation}
  \begin{split}
    \begin{aligned}
      p^\mu=p'^\mu=&(m,0,0,0)\\
      \epsilon^\mu=&(0,\vec\epsilon\,)\\
      \epsilon^{\ast\mu}=&(0,\vec\epsilon^{\,\ast})\\
      q^\mu=q'^\mu=&(\omega,\vec q\,),
    \end{aligned}
  \end{split}
  \label{useful}
\end{equation}
where $q$ and $q'$ are the 4-momenta of the incoming and 
outgoing photons, $\epsilon$ and $\epsilon^\ast$ their respective 
polarizations, while $p$ and $p'$ are the momenta of the 
incoming and ougoing baryons, respectively. 
We work in the Weyl gauge, which leads to the condition $p\cdot\epsilon=0$.

All terms containing the expression $\slashed\epsilon^\ast\slashed\epsilon$ 
contribute to $\gamma_0$, as can be seen when comparing Eq.~\ref{eqgamma0} 
with
\begin{align}
  \slashed\epsilon^\ast\slashed\epsilon=
  \mathrm{i}\vec{\sigma}(\vec\epsilon\times\vec{\epsilon}~^\ast)
  -(\vec{\epsilon}~\vec\epsilon~^\ast\,).
  \label{slesle}
\end{align}
Terms like $\slashed{\epsilon}^\ast\slashed q\slashed\epsilon$ yield a 
contribution of $-\mathrm{i}\omega\vec\sigma(\vec\epsilon\times\vec\epsilon~^\ast)$ 
when projected onto the baryon states. All the other expressions that arise can 
be reduced to this simple case.

While the full set of diagrams in the spin-$1/2$ sector is gauge invariant, special care 
has to be taken when including the spin-$3/2$ states. The diagrams which include a minimal 
coupling of the photon to the $\Delta(1232)$ would need 
terms of higher order to fully restore gauge invariance. In order to be fully gauge invariant the Lagrangian 
of Eq.~\ref{EqDelPhiB} should include the
covariant derivative $D_\mu\Delta_\nu$ as opposed to 
the partial derivative $\partial_\mu\Delta_\nu$ only. The difference between the two derivatives are 
higher-order terms. To solve this discrepancy, we follow the solution of 
Ref.~\cite{Lensky:2009uv}, where this problem has already been addressed for the case of the proton. In fact, 
for the neutral octet baryons the diagrams of Fig.~\ref{fiso32} are fully gauge invariant. As for the charged octet 
baryons, this is only the case for the diagrams with charged mesons. Therefore, for these baryons, the strategy is to 
study two sets of diagrams separately: on the one hand, we have the one-particle-irreducible diagrams of Figs.~\ref{fdela}, 
\ref{fdelg} and \ref{fdelh}, which are calculated summing over all isospin channels; on the other hand, the 
missing one-particle-reducible loop diagrams of Figs.~\ref{fdelb} to \ref{fdelf} are first calculated only for the charged 
meson channels. For the other channels the isospin factor is chosen such that the ratio between the isospins of the one-particle-reducible
and one-particle-irreducible diagrams is the same as for the charged meson channels. When doing so, gauge invariance 
is insured and the restoration procedure involves higher-order terms.

\section{Results and discussion}\label{sres}

The numerical results for our calculations, when including nucleons, 
pions and $\Delta$ resonances only (hadrons with no strangeness) are given 
in Table~\ref{tabsu2ours}, where a comparison with the numerical values found 
by other groups is also given. 
Our calculation for the isospin-$1/2$ sector completely agrees with the results of 
Ref.~\cite{Bernard:2012hb}. 
For completion, we also included how the $\gamma_0$ values vary when taking
the chiral limit, where the masses were set to the best-fit chiral masses. 
We compare our results with the HBChPT results 
from Refs.~\cite{Bernard:1992qa,Kao:2002cp}. The discrepancy between the results 
does not lie in the parameter choice but in the heavy-mass expansion one assumes 
for HBChPT.

As for the isospin-$3/2$ sector, our results differ from those of 
Ref.~\cite{Bernard:2012hb}. The reason for this is that we use a different counting scheme, and
therefore a different set of diagrams. We also have a different Lagrangian, which directly 
sorts out the spurious spin-$1/2$ contributions of the Rarita-Schwinger spin-$3/2$ spinor. In 
Ref.~\cite{Lensky:2014efa} the $\Delta(1232)$ was introduced in the same way as in the present work. 
One should remark that there a tree-level diagram of order $p^{9/2}$ was included, 
which we left out here for consistency. Without this diagram, 
the numerical results in Ref.~\cite{Lensky:2014efa} are in perfect agreement with ours. The decomposition of our results for the 
nucleon polarizabilities of Table~\ref{tabsu2ours} into their individual parts is listed in Table~\ref{tabdecomposition}. 
The main correction to the polarizability results comes from the tree-level diagrams with virtual spin-$3/2$ 
baryons, while their loop diagrams  give only a small contribution.

\begin{table}[h]
  \begin{center}
    \begin{tabular}{cc|ccccc|ccccc}
      \hline \hline
       \multicolumn{2}{c|}{\multirow{2}{*}{Model}}&\multicolumn{5}{c|}{proton}&\multicolumn{5}{c}{neutron}\\
                       && this work & \cite{Bernard:2012hb}& \cite{Bernard:1992qa}&\cite{Kao:2002cp}&\cite{Lensky:2014efa}& this work&\cite{Bernard:2012hb} & \cite{Bernard:1992qa}&\cite{Kao:2002cp}&\cite{Lensky:2014efa}\\
	  \hline
      \multirow{3}{*}{without $\Delta$}& HBChPT   &     &&4.4&& & & &4.4&\\
      &covariant chiral limit    & 2.15 & & & & &3.24 & &&&\\
      &covariant physical values & 2.07  & 2.07&& && 3.06 & 3.06&& &\\
	  \hline
      \multirow{3}{*}{with $\Delta$}& HBChPT    &   && &1.7& & & &&1.7&\\
      &covariant chiral limit    & -1.59(38) & & &&& -0.59(38) & &&&\\
      &covariant physical values & -0.76(28) & -1.74&& &-1.00 &0.15(28) &-0.77&&&  \\
	  \hline
	  \multicolumn{2}{c|}{experiment \cite{Hildebrandt:2003fm}} & \multicolumn{5}{c|}{$-1.01\pm0.08\text{(stat)}\pm0.10\text{(syst)}$}& &&&& \\
	  \multicolumn{2}{c|}{dispersion relations~\cite{Sandorfi:1994ku}} & \multicolumn{5}{c|}{$-1.34$} & \multicolumn{5}{c}{$-0.38$}\\
      \hline \hline
    \end{tabular}
  \end{center}
  \caption{Numerical values for $\gamma_0$ obtained in the SU(2) sector in present and in other works in units 
    of $10^{-4}$ fm$^4$. The choice of the numerical values for the constants for our own results
    can be found in Table~\ref{tabconsts}. The error in our results when including the $\Delta(1232)$ resonance 
arises from the uncertainty in the value of the low-energy constant $g_M$.
  }
  \label{tabsu2ours}
\end{table}
\begin{table}[h]
  \begin{center}
    \begin{tabular}{cc|c|c|c|c}
      \hline \hline
              &       & virtual spin-$1/2$ baryons & virtual spin-$3/2$ baryons --- tree level &  virtual spin-$3/2$ baryons --- loops  & total  \\\hline
       \multirow{2}{*}{$\gamma_0^p$}
	&   SU(2)              &       2.15&            -3.62		        &			 - 0.13          & -1.59  \\
       &SU(3)            &       1.53&            -3.62 		        &			 - 0.05          & -2.14  \\\hline
       \multirow{2}{*}{$\gamma_0^n$}
       &SU(2)&                 3.24		        &		  -3.62             &	 - 0.21          & -0.59  \\
       &SU(3) &       2.28	&            -3.62            	        &			 - 0.08          & -1.43  \\
       \hline \hline
    \end{tabular}
  \end{center}
  \caption{Decomposition of the proton and neutron polarizability results in units of $10^{-4}$ fm$^4$ into the contributions coming from the different sets of diagrams, when using the chiral limit for the masses and low-energy constants. The difference in results when using physical values or the chiral limit can be seen as systematical uncertainty.} 
  \label{tabdecomposition}
\end{table}

We extended the calculations to the SU(3) sector, again 
distinguishing between the case where the isospin-$3/2$ 
resonances were included and where only octet baryons were 
taken into account as intermediate states. Here we also considered 
both cases: when taking the physical average values and when choosing the chiral limit, 
see Table~\ref{tabconsts}. We also obtain predictions for the hyperon forward spin 
polarizabilities. A full listing of the results for the octet 
baryons is given in Table~\ref{tabsu3ours} and the decomposition of the results for the nucleons in Table~\ref{tabdecomposition}. 
When extending the model from SU(2) to SU(3), 
one takes into account additional virtual states and different values for the parameters, 
whose impact we discuss in more detail below. Another interesting feature in SU(3) is also 
the appearance of the SU(3)-flavour forbidden photon transitions of the negatively 
charged octet baryons to those of the decuplet. 
The results of Table~\ref{tabsu3ours} are 
also compared to HBChPT results 
(for preliminary results see Ref.~\cite{VijayaKumar:2011uw} and for a 
complete and improved analysis see Ref.~\cite{Astrid_Diplom}). 
For HBChPT the nucleon values for $\gamma_0$ change only slightly when going from SU(2) to SU(3), 
the results remain large and positive. When changing to the covariant version (without the decuplet 
contribution) the SU(3) case leads to a reduction of the $\gamma_0$ results which still 
are positive. An additional inclusion of the decuplet leads in both SU(2) and SU(3) cases to negative 
$\gamma_0$ values closer to the empirical value for the proton of 
$\left(-1.01\pm0.08\text{(stat)}\pm0.10\text{(syst)}\right)\cdot 10^{-4}$fm$^4$
presented in 
Ref.~\cite{Hildebrandt:2003fm}.

\renewcommand{\arraystretch}{2}
\begin{table}[h]
  \begin{center}
    \begin{tabular}{cc*{8}{c}}
      \hline \hline
      Model& used values&$p$&$n$&$\Sigma^+$&$\Sigma^-$&$\Sigma^0$
      &$\Lambda$&$\Xi^-$&$\Xi^0$\\
      \hline
      without decuplet, HBChPT & \cite{VijayaKumar:2011uw,Astrid_Diplom}&4.69&4.53&2.77&2.54&2.44&2.62&0.52&0.68\\
      \hline
      \multirow{2}{*}{without decuplet, covariant}& chiral limit&1.53&2.28&0.90&0.89&1.60&1.09&0.08&0.15\\
      & physical values&1.68&2.33&0.93&0.91&1.32&1.28&0.15&0.25\\
      \hline
      \multirow{2}{*}{with decuplet, covariant} & chiral limit&-2.14(38)&-1.43(33)&-2.72(33)&0.89&0.67(9)&-1.69(28)&0.07&-3.51(38)\\
      &physical values&-1.64(33)&-1.03(33)&-2.30(33)&0.90&0.47(8)&-1.25(25)&0.13&-3.02(33)\\
      \hline \hline
    \end{tabular}
  \end{center}
  \caption{Numerical values for $\gamma_0$ obtained in our calculations, in units 
    of $10^{-4}$ fm$^4$ in the SU(3) sector. The choice of the numerical values 
    for the constants in the covariant case can be found in Table~\ref{tabconsts}, 
    both for the chiral limit and for the physical average case. As for the HBChPT limit, 
    we cite the results in Ref.~\cite{VijayaKumar:2011uw}, which were later corrected 
    in our work in Ref.~\cite{Astrid_Diplom}. The errors in the results with the decuplet  
arise from the uncertainty in the value of the low-energy constant $g_M$.}
  \label{tabsu3ours}
\end{table}

It is also interesting to compare the $\gamma_0$ results for the nucleons to those 
from dispersion relation studies found in Ref.~\cite{Sandorfi:1994ku} to be 
$\gamma_0^p=-1.34\cdot 10^{-4}$fm$^4$ and $\gamma_0^n=-0.38\cdot 10^{-4}$fm$^4$. 
The inclusion of isospin-$3/2$ states, 
while already having an important effect in HBChPT, 
leads to an even better agreement with the empirical values in the case of fully 
covariant calculations, both when taking the chiral limit as well as 
when taking the average of the physical values for the constants. In fact, the 
difference between these two parameter sets is of higher chiral order for the 
polarizabilities. 
The main source of uncertainty of our results is 
the constant $g_M$, whose variation leads to an error estimate as shown in Table~\ref{tabsu3ours}.
We would like to stress that the results obtained here are not subject to 
uncertainties related to renormalization schemes; for the considered order there 
are no divergences or power-counting breaking terms entering into the value of $\gamma_0$.

As already discussed above, the inclusion of virtual decuplet states 
is crucial for an agreement of the nucleon B$\chi$PT polarizabilities with phenomenological values, 
which is dominantly because of the tree-level diagrams, as shown in Table~\ref{tabdecomposition}. 
However, this is not the case for the $\Sigma^-$ and $\Xi^-$ baryons since the photon transitions 
to the corresponding decuplet states $\Sigma^{*-}$ and $\Xi^{*-}$ are forbidden in SU(3) symmetry, 
and the tree-level diagrams do not appear. Hence, 
their values for the polarizabilities change only slightly by the small loop contributions with virtual 
decuplet baryons. 
To study the polarizabilites in B$\chi$PT, these two baryons might therefore be better suited 
than the proton and neutron since nearly all of the uncertainties coming from the inclusion 
of the decuplet drop out. Experimentally, it will be very hard to measure their polarizabilities, 
but it is feasable in lattice QCD. Furthermore, we want to emphasize 
that the main differences in numerical values for the proton and neutron polarizabilities in SU(2) 
and SU(3) come 
from the choice of the parameters in Table~\ref{tabconsts}. All contributions coming from $K$ 
or $\eta$ mesons are negligible. Choosing 
the masses and constants in SU(3) as in SU(2), which are equivalent parameter sets in 
terms of chiral counting up to the order $p^3$, will give nearly the same results.

The addition of $p^4$ contributions would be the next step to 
further refine the calculation.

\section{Summary}\label{summ}

We have presented an extended calculation for the spin polarizability $\gamma_0$
of the baryon octet. The framework we choose is based on manifestly Lorentz covariant
baryon chiral perturbation theory, both in the SU(2) as well as SU(3) versions. Furthermore, 
we explicitly include intermediate spin 3/2 states. 
The novel results of the present work concern both the SU(3) extension
of fully covariant ChPT to the order $p^3$ and the inclusion of the spin 3/2 decuplet
up to order $p^{7/2}$. Empirical results exist only for the nucleon case, where 
in both versions the inclusion of explicit decuplet states is crucial to find an agreement 
between phenomenology and B$\chi$PT. In particular, it is the tree-level 
diagram with virtual decuplet baryons that gives the dominant extra contribution. 
This also carries over to the SU(3) case, where all contributions from $K$ and $\eta$ loops 
turn out to be negligible. However, in SU(3) the two baryons $\Sigma^-$ and $\Xi^-$ are
 prominent as their photon transitions to the corresponding decuplet states are forbidden 
in SU(3) symmetry. As a result, the decuplet tree-level contributions are not present 
and their polarizabilities in pure B$\chi$PT remain nearly unchanged, meaning 
that also most of the uncertainties connected to the decuplet 
inclusion drop out. Since experimental polarizability measurements for 
these baryons are unprobable, comparisons to results from lattice QCD would 
be very interesting, as polarizabilities to chiral order $p^3$ are leading order predictions 
of B$\chi$PT. The $\gamma_0$ results for the hyperons, especially the ones 
for $\Sigma^-$ and $\Xi^-$, can therefore serve as a benchmark for
other calculations in this sector.
At this point it seems a necessity to extend the present calculation to the cases
of the electric and magnetic polarizabilities $\alpha_E$ and $\beta_M$ of the nucleon
and the baryon octet, especially the $\Sigma^-$ and $\Xi^-$, 
where probably a similar situation as above occurs with respect to the inclusion of decuplet states.

\clearpage
\subsection*{Acknowledgements}

This work was supported by Tomsk State University Competitiveness Improvement Program, 
by the Russian Federation program ``Nauka'' (Contract No. 0.1526.2015, 3854), 
by the Spanish Ministerio de Econom\'ia y Competitividad and European FEDER funds under Contracts No. FIS2011-28853-C02-01 and FIS2014-51948-C2-2-P, by Generalitat Valenciana under Contract No. PROMETEO/20090090 and by the EU HadronPhysics3 project, Grant Agreement No. 283286. A.N. Hiller Blin acknowledges support from the Santiago Grisol\'ia program of the  Generalitat Valenciana.
\appendix
\appendixpage
\section{Basic notations of ChPT}
\label{Appendix:ChPT}

The octet matrices of pseudoscalar mesons $\phi$, photons $Q$ and baryons $B$ 
are given by 
\begin{equation} 
\label{matrixmes}
\begin{split}
\begin{aligned}
\phi=\sum_{a=1}^8 {\lambda_a\phi^a}
&=\sqrt2\left(\begin{array}{ccc}
\frac 1{\sqrt 2}\pi^0+\frac1{\sqrt 6}\eta& \pi^+ &K^+\\
\pi^-& -\frac1{\sqrt 2}\pi^0+\frac1{\sqrt6}\eta& K^0\\
K^-&\bar{K}^0&-\frac 2 {\sqrt6}\eta
\end{array}\right),
\end{aligned}
\end{split}
\end{equation}

\begin{equation} 
\label{photonmes}
\begin{split}
\begin{aligned}
Q=\frac{1}{2}\left(\lambda_3+\frac{\lambda_8}{\sqrt{3}}\right)
&=\left(\begin{array}{ccc}
\frac 23& 0&0\\
0&-\frac13&0\\
0&0&-\frac13
\end{array}\right)
\end{aligned}
\end{split}
\end{equation}

and 
\begin{equation}
\begin{split}
\begin{aligned}
B=\frac1{\sqrt2}\sum_{a=1}^8 {\lambda_a B^a}
&=\left(\begin{array}{ccc}
\frac 1{\sqrt 2}\Sigma^0+\frac1{\sqrt 6}\Lambda& \Sigma^+ &p\\
\Sigma^-& -\frac1{\sqrt 2}\Sigma^0+\frac1{\sqrt6}\Lambda& n\\
\Xi^-&\Xi^0&-\frac 2 {\sqrt6}\Lambda
\end{array}\right).
\end{aligned}
\end{split}
\label{matrixbar}
\end{equation}

The decuplet states $\Delta^{ijk}$ are specified as 
\eq 
& &\Delta^{111}=\Delta^{++}\,, \ 
   \Delta^{112}=\frac1{\sqrt3}\Delta^{+}\,, \ 
   \Delta^{122}=\frac1{\sqrt3}\Delta^{0}\,, \ 
   \Delta^{222}=\Delta^{-}\,,\nonumber\\
& &\Delta^{113}=\frac1{\sqrt3}\Sigma^{\ast+}\,, \
   \Delta^{123}=\frac1{\sqrt6}\Sigma^{\ast 0}\,, \ 
   \Delta^{223}=\frac1{\sqrt3}\Sigma^{\ast -}\,, \nonumber\\
& &\Delta^{133}=\frac1{\sqrt3}\Xi^{\ast 0}\,, \ 
   \Delta^{233}=\frac1{\sqrt3}\Xi^{\ast -}\,, \
   \Delta^{333}=\Omega^{-}. \,
\en

The Dirac tensors $\gamma^{\mu\nu}$ are defined as 
\eq 
\gamma^{\mu\nu}
=\frac12\left[\gamma^\mu,\gamma^\nu\right]\hspace{.5cm}\text{and}\hspace{.5cm}
\gamma^{\mu\nu\lambda}=\frac14\left\{\left[\gamma^\mu,\gamma^\nu\right],
\gamma^\lambda\right\}\,.
\en 

\section{Loop integrals and dimensional regularization}\label{apints}

The integrals for Figs.~\ref{fiso12} and \ref{fiso32} are taken in $d$ dimensions and 
later dimensionally regularized to the normal 4-dimensional 
Minkowski space. 
To calculate 
the crossed diagrams, the simple substitutions
\begin{equation}
  \omega\leftrightarrow-\omega\hspace{5mm}
  \text{and}\hspace{5mm}\epsilon^*\leftrightarrow\epsilon
\end{equation}
have to be performed. To obtain the final numerical results for 
the forward spin polarizability, the integrands of the structure 
constants of the $\sle^*\sle$-terms are expanded up to the order 
$\omega^3$. The coefficients of 
the third order of the expansion are then used to evaluate the integrals.

The following loop integrals are of interest in this work:

\begin{align}
\nn \int{\frac{\mathrm{d}^dz}{(2\pi)^d}\frac{1}{\left(z^2-\Delta\right)^n}}
&=\frac{(-1)^n\mathrm{i}}{(4\pi)^{d/2}}
\frac{\Gamma\left(n-\frac d2\right)}{\Gamma(n)\Delta^{n-\frac d2}}\\
\nn \int{\frac{\mathrm{d}^dz}{(2\pi)^d}
\frac{z^\mu z^\nu}{\left(z^2-\Delta\right)^n}}
&=\frac{(-1)^{n-1}\mathrm{i}}{(4\pi)^{d/2}}
\frac{\Gamma\left(n-\frac d2-1\right)}{\Gamma(n)
\Delta^{n-\frac d2-1}}\frac {g^{\mu\nu}}{2}\\
\nn \int{\frac{\mathrm{d}^dz}{(2\pi)^d}
\frac{z^\mu z^\nu z^\rho z^\sigma}{\left(z^2-\Delta\right)^n}}
&=\frac{(-1)^{n}\mathrm{i}}{(4\pi)^{d/2}}
\frac{\Gamma\left(n-\frac d2-2\right)}{\Gamma(n)\Delta^{n-\frac d2-2}}
\frac {g^{\mu\nu}g^{\rho\sigma}+g^{\mu\rho}
g^{\nu\sigma}+g^{\mu\sigma}g^{\nu\rho}}{4}\,.
\end{align}
As a result, one has to dimensionally regularize the integrals, 
obtaining the expressions
\begin{align*}
 \lambda_1(\Delta)=
\frac{\Gamma\left(1-\frac d2\right)}{(4\pi)^{d/2}\Delta^{1-\frac d2}}
&= -\frac{\Delta}{16\pi^2}\left(\frac{2}{\epsilon}
-\log\left(\frac{\Delta}{\mu}\right)+\log(4\pi)
-\gamma_E+1+\mathcal{O}(\epsilon)\right)\\
\lambda_2(\Delta)=\frac{\Gamma\left(2-\frac d2\right)}
{(4\pi)^{d/2}\Delta^{2-\frac d2}}
&= \frac{1}{16\pi^2}\left(\frac{2}{\epsilon}
-\log\left(\frac{\Delta}{\mu}\right)+\log(4\pi)
-\gamma_E+\mathcal{O}(\epsilon)\right)\\
\lambda_3(\Delta)=\frac{\Gamma\left(3-\frac d2\right)}{(4\pi)^{d/2}
\Delta^{3-\frac d2}}
&= \frac{1}{16\pi^2\Delta}\\
\lambda_4(\Delta)=\frac{\Gamma
\left(4-\frac d2\right)}{(4\pi)^{d/2}\Delta^{4-\frac d2}}
&= \frac{1}{16\pi^2\Delta^2},
\end{align*}
where $\epsilon=4-d$ and $\mu$ is the scale parameter 
set to the proton mass in this work. For regularization, 
the minimal subtraction ($\widetilde{\text{MS}}$) would have to be performed 
in the EOMS scheme, where terms proportional to
\begin{align}
\nn \frac{2}{\epsilon} + \log(4\pi) - \gamma_E + 1
\end{align}
are subtracted.
 It is interesting to note that in this work 
no diagram had to be renormalized, as at order $p^{7/2}$ no divergent 
or power counting breaking terms contribute to $\gamma_0$.


\end{document}